\documentclass[aps,amsmath,superscriptaddress,twocolumn]{revtex4}
\usepackage{graphicx}

\begin{document}

\title{Message in the Sky}

\author{S.~Hsu} \email{hsu@duende.uoregon.edu}
\affiliation{Institute of Theoretical Science \\ University of Oregon,
Eugene, OR 97403}

\author{A.~Zee}
\email{zee@itp.ucsb.edu}
\affiliation{Kavli Institute for Theoretical Physics \\
University of California, Santa Barbara, CA 93106}

\begin{abstract}
We argue that the cosmic microwave background (CMB) provides a
stupendous opportunity for the Creator of universe our (assuming one
exists) to have sent a message to its occupants, using known physics.
Our work does not support the Intelligent Design movement in any way whatsoever, but asks, and attempts to answer, the entirely scientific question of what the medium and message might be IF there was actually a message.  The medium for the message is unique. We elaborate on this observation, noting that it requires only careful adjustment of the fundamental Lagrangian, but no direct intervention in the subsequent evolution of the universe.
\end{abstract}


\maketitle

\date{today}

Let us pose the following question. Suppose some superior Being or
Beings got the universe going. We do not address the issue of whether
or not this is likely, but merely proceed with this
supposition. Furthermore, suppose that they actually wanted to notify
us that the universe was intentionally created. The question we would
like to ask is: How would they send us a message?

That the universe was started by superior Beings is not only the
province of religious thoughts from the earliest days of the human
race, but has also been a staple of science fiction. In one of our
favorite scenarios, our universe is a school-assigned science
experiment \cite{fearful,bubble} carried out by a high school student
in a meta-universe. Perhaps he or she or it even started an assortment
of universes like ant farms and stashed them away somewhere in the
basement, out of his or her or its parent's way. Perhaps by now he has
lost interest and forgotten about the universes, leaving some to
expand, others to collapse, in complete futility and silence. But,
perhaps not without leaving a message for the occupants...

The popular press has been rife with suggestions on where the message
might be found. One suggestion, for example, is that there might be a
coded message in the human genome. In the United States, people with
certain religious convictions have even imagined that the message
might be encoded in the rock formation of the Grand Canyon, as another
example. In our opinion, such suggestions are clearly not universal
enough, and they seem to require direct intervention by the Creator
during the evolution of the universe.  Another possibility, that a
message might be hidden deep in the digits of pi or the Riemann zeta
function, is also appealing, but we have no way of addressing how
feasible that might be without some measure over possible realizations
of mathematics.

If one of the present authors had gotten the universe going and if he had
wanted to announce this fact, he would clearly want all the advanced
civilizations, not just in our galaxy, but in the entire universe, to know.
The genomic sequence is limited to civilizations involving
(presumably) some subset of possible life forms based on carbon. To think
that the message is carved in the geological record of one particular
country on some insignificant dust speck of a planet is somewhat far fetched
to say the least.

We have convinced ourselves that the medium for the message is unique:
it could only be the cosmic microwave background. The cosmic microwave
background is in effect a giant billboard in the sky, visible to all
technologically advanced civilizations. Since different regions of the
sky are causally disconnected, only the Being ``present at the
creation'' could place a message there. (There are also cosmic
neutrino and gravity wave backgrounds, but given the elusive nature of
neutrinos and gravity waves it seems that photons are a better choice
for carriers of the message.)

The variances of the CMB temperature distribution, projected onto spherical
coordinates, are given by numbers denoted by $C_{l}$ \cite{dodel}. Here $l$
is the usual angular momentum quantum number which appears in spherical
harmonics $Y_{lm}$. Currently planned experiments are capable of measuring $
C_{l}$'s up to $l\sim 10^{4}$. There is a fundamental (statistical) limit on
the accuracy $\Delta C_{l}$ to which we can determine each $C_{l}$, due to
so-called cosmic variance. Since we only have $(2l+1)$
possible samples, we obtain
\begin{equation}
\left( \frac{\Delta C_{l}}{C_{l}}\right) =~\sqrt{\frac{2}{2l+1}}.
\end{equation}
Aside from this cosmic variance, each observer in the universe sees
the same $C_l$'s, regardless of location. This point is worth
emphasizing: while different civilizations, due to their different
locations, see different patterns of thermal fluctuations on the sky
(since their past light cones sample different portions of the surface
of last scattering), they nevertheless would measure the same $C_l$'s
and agree on the probability distribution from which the different
patterns were drawn.

The next question is what might the message be. We thought of various
possibilities and decided that the best choice would be the following. We
now know, and we suppose that any civilization advanced enough to detect $%
C_{l}$ in the comic microwave background would also know, that three of the
four fundamental interactions are governed by gauge theories, based on the
Lie algebras $U(1)\otimes SU(2)\otimes SU(3)\subset SU(5)\subset
SO(10)\subset E(6)$ \cite{exceptional}. Thus, we suggest that the coded
message would simply be an announcement along the line ``Hey guys, the
universe is governed by gauge theories, and the relevant algebras are such
and such.''

As is well known, all Lie algebras are uniquely characterized \cite{slan} by
Dynkin diagrams or alternatively Cartan matrices. The Cartan matrices are
simply square matrices whose entries are mostly zero, with the non-zero
entries equal to either 2 or $-1$. For example the Cartan matrix for $E(6)$
is
\begin{equation}
\left(
\begin{array}{llllll}
2 & -1 & 0 & 0 & 0 & 0 \\
-1 & 2 & -1 & 0 & 0 & 0 \\
0 & -1 & 2 & -1 & 0 & -1 \\
0 & 0 & -1 & 2 & -1 & 0 \\
0 & 0 & 0 & -1 & 2 & 0 \\
0 & 0 & -1 & 0 & 0 & 2
\end{array}
\right)
\end{equation}
This could easily be coded as a sequence of bits, with the first bits
specifying the coding system (by listing the simplest algebras in
order of complexity), and later bits describing the actual gauge
``theory of not quite everything.''

We emphasize that the message is hidden in very small temperature
fluctuations in the CMB (of order $10^{-5}$), presumably resulting
from primordial density perturbations. As a specific suggestion for
how the encoding might be done, let us assume that inflationary
cosmology (to which we are by no means committed for the purpose of
this paper) is correct and suppose that before the Big Bang the
Creator fine tuned the inflaton potential with small deviations from
flatness. These would lead to deviations from the usual
scale-invariant spectrum of density perturbations.  The three symbols
($2, -1, 0$) appearing in the Cartan matrices can be encoded by
adjusting the potential to produce above, below or exactly average
amplitudes of fluctuation during the evolution of $\phi$.  (A ``full
stop'' or period might be signified by a sharper spike in amplitude.)
These amplitudes would be reflected by variances at different length
scales, or equivalently, values of $C_l$.  In the slow-roll
approximation, the amplitude of density fluctuations is given by the
ratio between the square of the Hubble parameter $H^2$ and the rate of
change of the inflaton field, $\dot{\phi}$, and so is controlled by
the precise shape of the inflaton potential \cite{slow}. An advanced
civilization, having independently measured the number of baryons,
amount of dark energy, and so forth in the universe, could detect any
small deviations from what would have resulted from scale invariant
density perturbations, and interpret these as an encoded signal.  We
also assume that (as one definition of ``advanced'') an advanced
civilization will be able to de-convolve subsequent evolution of the
CMB due to classical physics.

Regardless of the method employed, the amount of information that
could be encoded in the numbers $C_{l}$ is limited, due to cosmic
variance. Assuming $l<l_{\text{max}}$, how well can the $C_{l}$'s be
measured? If the limiting resolution is of order $1/\sqrt{l}$, then
the logarithm of the total number of possible messages is of order
\begin{equation}
\ln N\sim \ln \prod_{l=1}^{l_{\text{max}}}\sqrt{l}\sim \frac{1}{2}%
\sum_{l=1}^{l_{\text{max}}}\ln l\sim \frac{1}{2}l_{\text{max}}\ln l_{\text{
max}}~.
\end{equation}
Taking $l_{\text{max}}\ \sim 10^{4}$, we estimate no more that
$10^{5}$ bits of information can be encoded. This would be easily enough to
give the first few dozen or so Cartan matrices (to establish the
pattern of encoding), the final result specifying the grand-unified
gauge group and perhaps the matter representations.

Of course this is just one specific suggestion. Perhaps our collective
scientific mind is still too puny to guess what the message on the billboard
in the sky might read. For instance, string theorists might argue that the
message would simply specify the correct string theory, assuming that such a
thing exists. Certainly, many other possible messages exist \cite{prime}.

We can also think of other ways of encoding \cite{fundconst} the
message itself, and again our present knowledge of quantum mechanics
and cosmology may be insufficiently advanced.  Our current
understanding is that the primordial fluctuations are due to quantum
fluctuations. Perhaps the superior Beings are able to manipulate
quantum fluctuations at will. (That may be taken as one manifestation
of being superior.) We simply do not understand quantum mechanics
beyond the algorithmic level. Or perhaps the ``finger of God'' came
down during recombination and photon decoupling to imprint the desired
fluctuations on the cosmic microwave background. While requiring
active intervention in our universe, this is at least far more
plausible than the finger of God rearranging the genomic sequence of
Homo sapiens at some point during our evolution.

Coming to the end of our short paper, we allow ourselves to indulge in
a wild speculation. Even if a hidden message turns up in the CMB,
nothing in this paper requires that the superior Being (or the
Ultimate Designer \cite{fearful} as one of us prefers) is theistic
rather than deistic in character \cite{thedei}. On the other hand,
perhaps the Ultimate Designer --- the pimply-faced teenager mentioned
in the introduction --- could himself, herself, or itself live in a
theistic universe. There may be levels of universes, with Ultimate
Designers all the way up. Even if our own universe, which we now
understand to occupy the lowest level, may be merely deistic, some of
these universes could well be theistic. We could easily persuade
ourselves that the notion of a hierarchy of universes is more
appealing than the notion of casually disconnected parallel universes.

In conclusion, we believe that we have raised an intriguing
possibility: a universal message might be encoded in the cosmic
background. When more accurate CMB data becomes available, we urge
that it be analyzed carefully for possible patterns.  This may be even
more fun than SETI.

\bigskip
{\bf Note added}: Subsequent to this paper, another, much smaller, estimate of the observer-independent information in the CMB was given by Scott and Zibin in arXiv:physics/0511135. In contrast to the upper bound of $10^5$ bits deduced here, Scott and Zibin's estimate requires additional assumptions about the dynamic range of the spectrum of density perturbations produced by inflation. In particular, their estimate assumes that the resulting range of variation in the $C_l$'s is no greater than a few percent (roughly one unit of cosmic variance, for $l \sim 1000$) over a change in $l$ of about 10. This assumption seems to us rather arbitrary, and has not been justified from first principles. Nevertheless, their smaller estimate of 300 bits of information would be enough to encode the simplest Cartan matrices (as a key) and the GUT breaking pattern.

\bigskip

\textbf{\large Acknowledgments\medskip }

SH thanks A.~Kosowsky and B.~Murray for useful comments. AZ would like
to thank the Institute for Theoretical Sciences at the University of
Oregon, where this work was initiated. SH is supported by the
Department of Energy under DE-FG06-85ER40224. AZ is supported in part
by the National Science Foundation under grant PHY04-56556.

\bigskip



\begin{thebibliography}{9}
\bibitem{exceptional}  It is an intriguing fact that the sequence $
E(6)\rightarrow SO(10)\rightarrow SU(5)$ could be obtained by sucessively
chopping off a circle starting with the Dynkin diagram for $E(6).$
Mathematically, $SO(10)$ could be regarded as $E(5)$ and $SU(5)$ as $E(4).$
\bibitem{slow} To encode many symbols might require rapid variation of
$\phi$, violating the slow-roll approximation. But of course our
current understanding of inflationary dynamics is quite rudimentary --
limited to models which are easy to analyze.
\bibitem{prime}  For example, another suggestion might be the sequence of
prime numbers, but this strikes us as not informative enough. (One may even
conceive of civilizations for which the prime numbers may not hold as much
fascination as for our own.)
\bibitem{fundconst} Another possibility might be to hide the message
inside the fine structure ``constant'' or some other parameter of
physics. Aside from the fact that the fine structure constant evolves
due to renormalization (hence is scale dependent), experimental
considerations may make it hard to measure any constant to $10^{5}$
bits of accuracy. Consequently, the encoded message must be even
shorter.
\end{thebibliography}
\end{document}